\documentclass[]{iopart}
\usepackage{graphicx}
\newcommand{\sn}{\rm sn}
\newcommand{\cn}{\rm cn}
\newcommand{\dn}{\rm dn}
\newcommand{\sca}{\rm sc}
\newcommand{\sech}{\rm sech}
\newcommand{\be}{\begin{equation}}
\newcommand{\ee}{\end{equation}}
\newcommand{\bea}{\begin{eqnarray}}
\newcommand{\eea}{\end{eqnarray}}

\newcommand{\thirdfig}{.3\textwidth}
\newcommand{\middlefig}{.5\textwidth}

\begin{document}

\title{On Some Classes of mKdV Periodic Solutions}

\author{P.G. Kevrekidis$^1$, Avinash Khare$^2$, A. Saxena$^3$ and 
G. Herring$^1$}
\address{$^1$ Department of Mathematics and Statistics, University
of Massachusetts, Amherst, MA 01003-4515, USA.}
\address{$^2$ Institute of Physics, 
Bhubaneswar, Orissa 751005, India.}
\address{$^3$ Theoretical Division, Los 
Alamos National Laboratory, Los Alamos, New Mexico 87545, USA.}

 \begin{abstract}
We obtain exact  periodic  
solutions of the  positive and  negative modified 
Kortweg-de Vries (mKdV) equations.   
We examine the dynamical stability of these solitary wave
lattices through direct numerical simulations. While the positive
mKdV breather lattice solutions are found to be unstable, the
two-soliton lattice solution of the same equation is found to be
stable. Similarly, a negative
mKdV lattice solution is found to be stable. We also 
touch upon the implications of these results for the KdV equation. 

\end{abstract}

\submitto{\JPA}
\maketitle

\section{Introduction}
Among the soliton bearing nonlinear integrable equations, sine-Gordon, 
nonlinear Schr{\"o}dinger (NLS), 
Korteweg-de Vries (KdV) and 
the modified Korteweg-de Vries (mKdV) are of special interest 
\cite{mkdv1,AS,IR,Dodd,Scott}.  
These equations possess exact 
breather solutions.  Therefore, they may also have exact solutions 
in the form of a spatially periodic array of single breathers, i.e. breather 
lattices.  Similarly 
to the sine-Gordon, NLS and KdV, the mKdV equation is also 
important in many physical contexts. For example, it appears in the context
of ion acoustic solitons \cite{ion}, van Alfv\'en waves in collisionless
plasma \cite{plasma}, Schottky barrier transmission lines \cite{schot}, 
models of traffic congestion \cite{traffic} as well as 
phonons in anharmonic lattices \cite{ono}. Furthermore, the modelling of 
a subclass of hyperbolic surfaces \cite{schief}, slag-metallic 
bath interfaces \cite{slag} as well as meandering 
ocean jets \cite{jets} is also related to the mKdV equation. 
The dynamics of thin elastic rods has also been demonstrated to be
reducible to the mKdV
equation \cite{mat}. Finally, if one studies the examples of surface dynamics
that are purely local, yet maintain global constraints like
conservation of perimeter and enclosed area, one finds that these dynamics are
closely related to the KdV and mKdV hierarchies \cite{gol}.  

The nonlinear term in the mKdV equation 
($6u^2u_x$) may assume either a positive or a negative sign. We will
classify the equation as ``positive mKdV'' and ``negative mKdV''
if the prefactor is $+1$ or $-1$, respectively. In an
earlier study,  
the stability of the sine-Gordon breather lattice was examined 
\cite{sg}.  Recently, a particular  form of an 
exact breather lattice solution was obtained in \cite{mkdv} for
the positive mKdV equation and its stability was discussed.

The aim of the present Letter is to present a new class of periodic 
solutions {\it both} for the positive and for
the negative mKdV equations. We also intend to examine the dynamical
stability of these novel classes of solutions and particularly to
illustrate that many of them {\it can be dynamically stable}. This
is notably different from the behavior observed previously 
for breather lattice solutions in the models of \cite{sg,mkdv}.
Furthermore, it is worth noting that the solutions of the negative 
mKdV are related to those of the KdV equation via the Miura transform 
\cite{miura}.  Thus, our solutions can be translated into 
exact periodic solutions of 
the KdV equation as well. The latter is also a ubiquitous equation
in a variety of fields
ranging from conformal field theory to plasma physics.


Our presentation is structured as follows. In Sec. 2 we examine the 
breather lattice and two-soliton lattice solutions of the mKdV equation with 
positive sign of nonlinearity and  discuss their stability. In Sec. 3 we 
examine periodic solutions of the mKdV equation
 with the negative sign 
of the nonlinearity. In Sec. 4 we briefly comment on the corresponding 
KdV solutions and follow that with our conclusions in Sec. 5. 

\section{Breather and Soliton Lattices of the Positive mKdV Equation} 

For a field $u(x,t)$, the positive mKdV equation is given by 
\begin{equation}
u_t+6u^2 u_x+u_{xxx} =0.
\label{aveq1}
\end{equation}
Using the ansatz
\begin{equation}
u=-2\frac{d}{dx} [\tan^{-1} \phi], 
\label{aveq2}
\end{equation}
we are able to
obtain several exact spatially periodic solutions.  The first one is a 
breather lattice solution \cite{mkdv} 
\begin{equation}
\phi = \alpha \sn(ax+bt+a_0,k)\dn(cx+dt+c_0,m),
\label{aveq3}
\end{equation}
where $a_0,c_0$ are arbitrary constants. In fact, all the solutions 
discussed in this paper admit such constants even though we will not 
always display them hereafter. 
In this solution: 
\begin{eqnarray}
\alpha = -\frac{c}{a},~ \frac{c^4}{a^4}=\frac{k}{(1-m)},
~ \frac{b}{a}=[a^2(1+k)-3c^2(2-m)], \nonumber\\ 
\frac{d}{c}=[3a^2(1+k)-c^2(2-m)]\,,
\label{aveq4}
\end{eqnarray}
while \sn$(x,k)$, ($\cn(x,k)$ below), and $\dn(x,m)$ are Jacobi elliptic 
functions with modulus $k$ and $m$, respectively.  This solution was 
studied in detail previously in \cite{mkdv}. For completeness, we 
discuss briefly the relevant results. In order to ensure periodicity
of the solution (in space), a commensurability condition was postulated
in that work. In its strict form (strong commensurability), this condition 
demands that the two elliptic function terms of Eq. (\ref{aveq3})
have the same period, i.e. $4 K(k)/a=2 K(m)/c$ (where $K(k)$ 
denotes the complete elliptic integral of the first kind). In its weaker
form (weak commensurability), the condition demands that
the periodicities are rational multiples of each other i.e.,
\begin{equation}
4 p \frac{K(k)}{a} = 2 q \frac{K(m)}{c}
\label{aveq5}
\end{equation}
with $p,q \in Z_+^{\star}$, note that $p=q=1$ yields the strong commensurability
as a special case. Both strong and weak forms, however, resulted in 
{\it unstable} dynamical evolutions of the breather lattice \cite{mkdv}  
(induced by means of numerical perturbations to the exact solution).
We further showed that this solution could be stabilized through ac driving
and damping. In the limit $k \rightarrow 0,m \rightarrow 1,$ but with 
${k}/(1-m) = v^4$, this solution goes over to the well known single 
breather (bion) solution
\begin{equation}
\phi = -v \sin(ax+bt+a_0) \sech (cx+dt+c_0).
\label{aveq6}
\end{equation}

\begin{figure}
\begin{center}
\begin{tabular}{ccc}
    \includegraphics[width=\thirdfig]{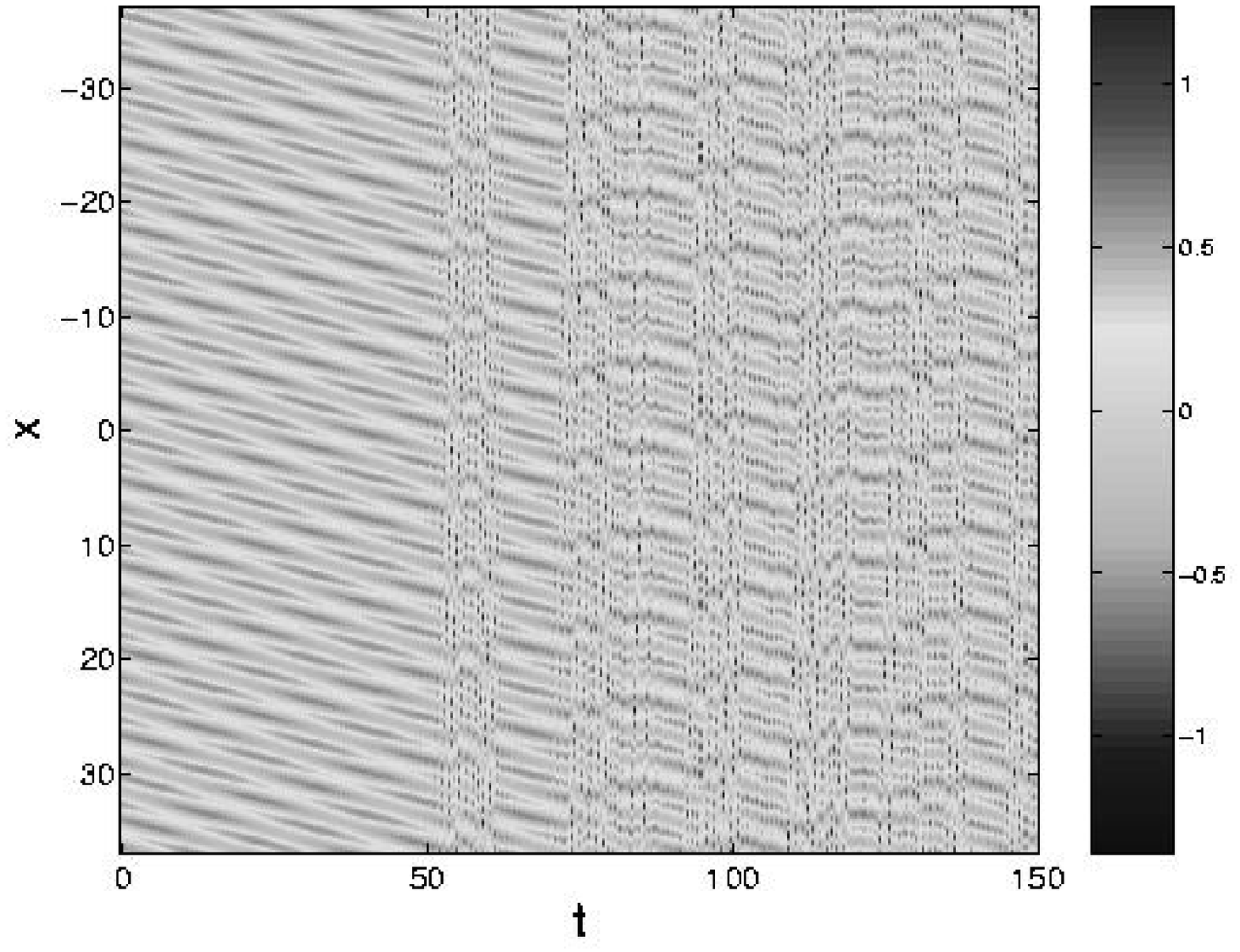}&
    \includegraphics[width=\thirdfig]{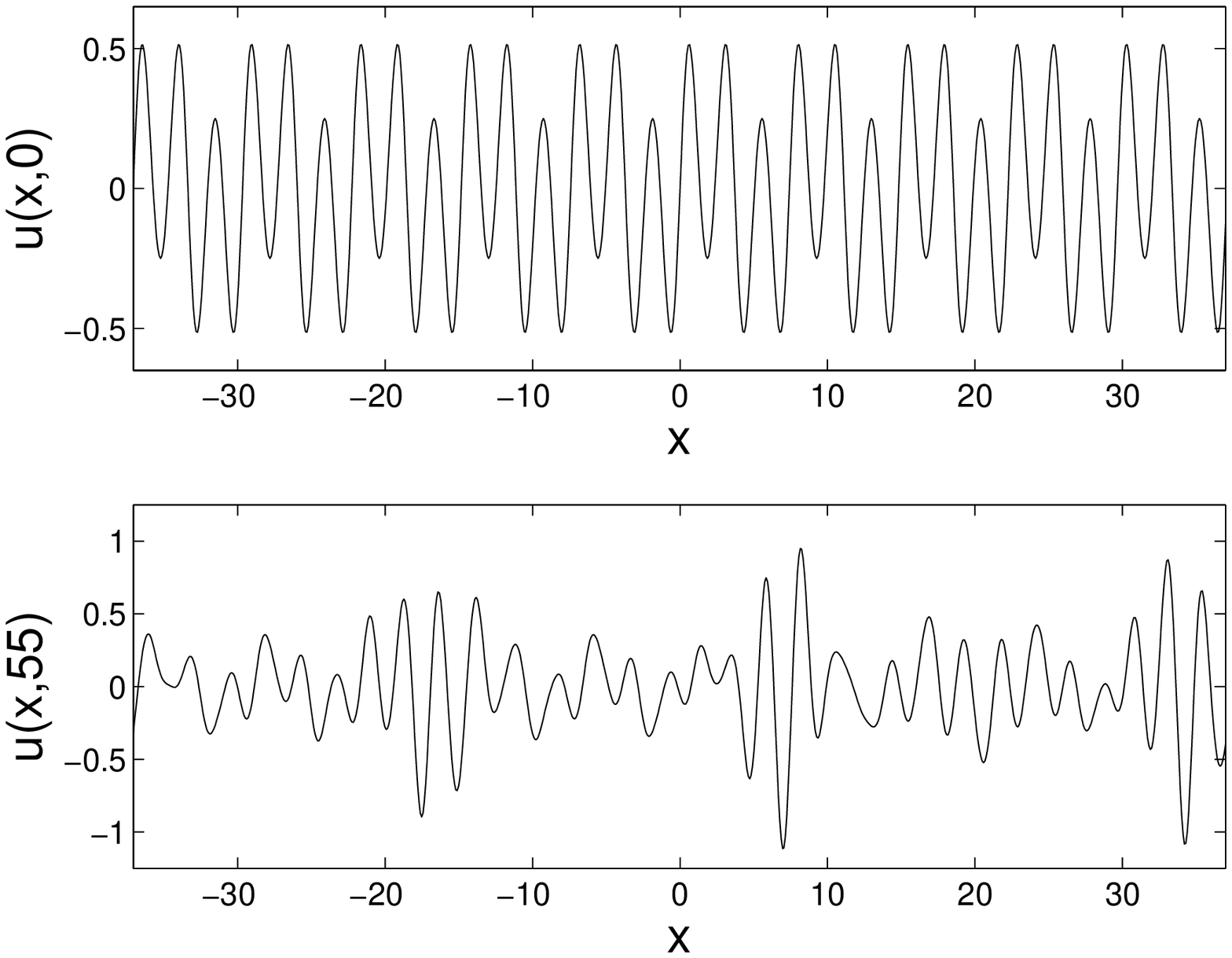}&
    \includegraphics[width=\thirdfig]{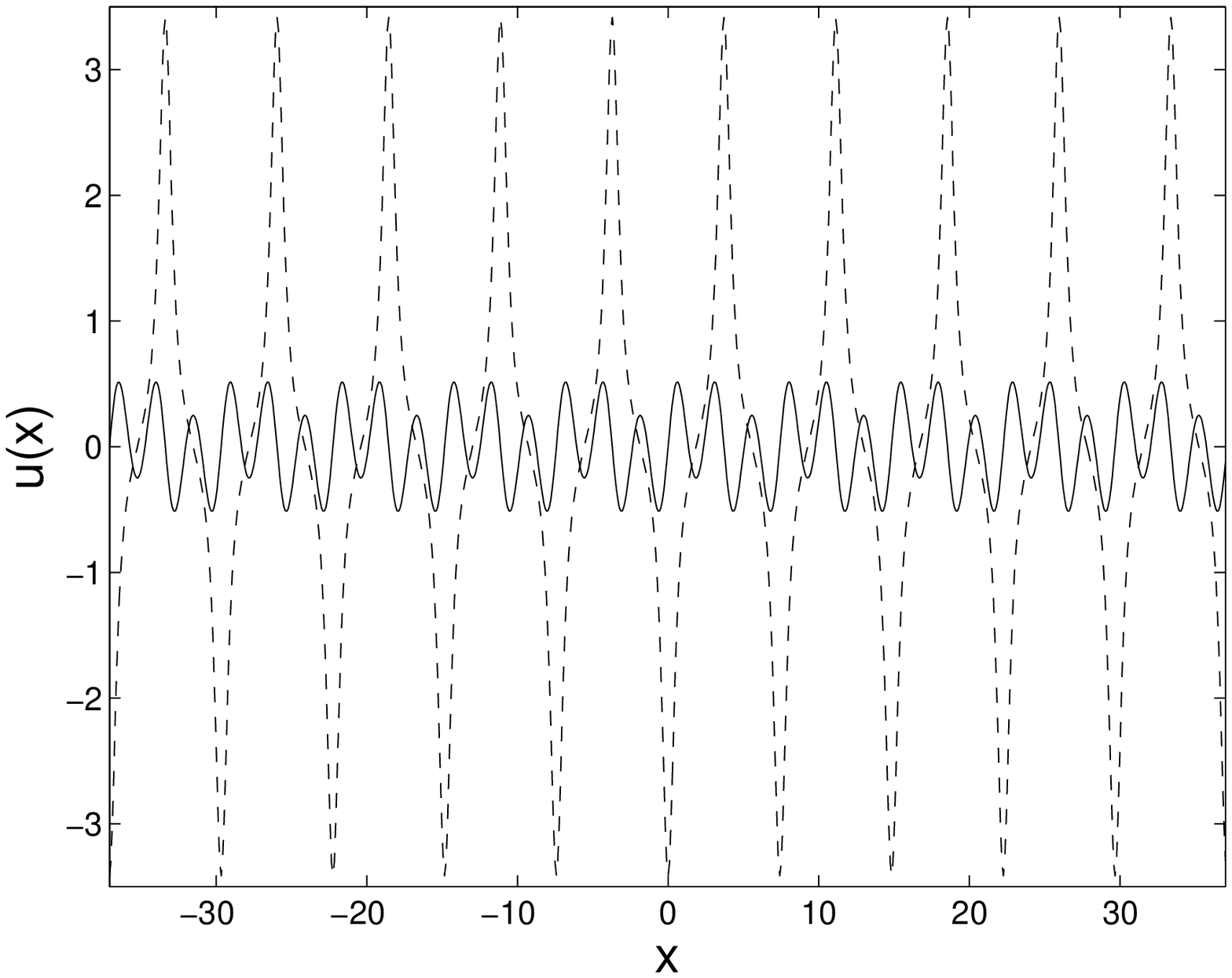}\\
\end{tabular}
\caption{The left panel shows the x-t (i.e., space-time) evolution
of the contour of the unstable breather lattice solution of the
form of Eqs. (\ref{aveq7})-(\ref{aveq8}); $a=1$, $k=0.5$, $c=1.707$
and $m=0.03$ (such that $K(k)/a=2 K(m)/c$) were 
used in the initial conditions. 
The middle panel shows the spatial profile at $t=0$ (initial
condition) and at $t=55$ (after the instability has set in). 
Finally, the right panel
shows with a solid line the initial condition and with a dashed line
the (other breather lattice) 
solution of Eqs. (\ref{aveq3})-(\ref{aveq4}) for the same parameters.}%
\label{avfig1}
\end{center}
\end{figure}

Remarkably, it turns out that there is a different breather lattice solution to
the positive mKdV equation 
given by
\begin{equation}
\phi = \alpha \cn(ax+bt+a_1,k)\cn(cx+dt+c_1,m),
\label{aveq7}
\end{equation}
where 
\begin{eqnarray}
\alpha^2 = \frac{mk}{(1-m)(1-k)},~\frac{c^4}{a^4} 
= \frac{k(1-k)}{m(1-m)}, \nonumber \\
\frac{b}{a}=[a^2(1-2k)+3c^2(1-2m)],~\frac{d}{c}=[3a^2(1-2k)+c^2(1-2m)].
\label{aveq8}
\end{eqnarray}
Notice that in the limit $k \rightarrow 0,m \rightarrow 1, a_1 =\pi /2+a_0$
$c_1 = c_0$ but with ${k}/(1-m) =v^4$, this solution also goes over to the same
breather (bion) solution (\ref{aveq6}). However, for other values of $k,m$,
the two breather lattice solutions (\ref{aveq3}) and (\ref{aveq7}) 
are quite different (see also the comparison of the right panel
of Fig. \ref{avfig1}). 
We should note here that the characterization of the solutions that
we present as breather or two-soliton lattices is given on the basis
of what their limiting (see e.g., Eq. (\ref{aveq6})) profile looks like, 
as the elliptic functions asymptote to trigonometric/hyperbolic ones.

The dynamical stability of the breather lattice solution (\ref{aveq7}) 
was examined for 
both strong and weak commensurability by means of direct
numerical simulations. The numerical scheme used
 here, motivated
by the KdV discretization of Ref. \cite{ohta} was analyzed previously 
in \cite{mkdv}. 
However, the results were also verified with different discretizations
of the integrable model such as the ones proposed in \cite{ladik}.
In Fig. \ref{avfig1}, we demonstrate a typical example of the dynamical
evolution of the solution of Eqs. (\ref{aveq7})-(\ref{aveq8}). We observe
that similarly to the previously obtained breather lattice 
solution of \cite{mkdv},
the breather lattice identified above is dynamically unstable in the
mKdV equation and results in a few more strongly and many weakly localized
peaks. Hence, the instability of mKdV breather lattices appears to be 
generic.

Using the ansatz in terms of Eq. (\ref{aveq2}), we can obtain yet another
novel family of solutions, namely a two-soliton lattice 
\begin{equation}
\phi = \alpha \sca (ax+bt,k)\dn (cx+dt,m),
\label{aveq9}
\end{equation}
where $\alpha=-c/a$, $\sca(x,m) \equiv \sn(x,m)/\cn(x,m)$ and 
\begin{eqnarray}
 \frac{c^4}{a^4}=\frac{1-k}{1-m}, ~~\frac{b}{a}
=-[a^2(2-k)+3c^2(2-m)], \nonumber\\ 
\frac{d}{c}=-[c^2(2-m)+3a^2(2-k)].
\label{aveq10}
\end{eqnarray}
This solution can also be obtained from the breather lattice solution 
(\ref{aveq3}) by taking $a \rightarrow ia,b \rightarrow ib, \alpha \rightarrow
-i\alpha$ and using the well known relations
\bea\label{aveq11}
&&\sn(ix,m) = i\frac{\sn(x,1-m)}{\cn(x,1-m)}\,,~~
\dn(ix,m) = \frac{\dn(x,1-m)}{\cn(x,1-m)}\,, \nonumber \\
&&\cn(ix,m) = \frac{1}{\cn(x,1-m)}\,.
\eea
In the limit $k \rightarrow 1,\,m \rightarrow 1$ but with $(1-k)/(1-m)=v^4$
this solution goes over to 
\be\label{aveq12}
\phi = -v\sinh[a(x-[1+3v^2]t)]\sech[av(x-[3+v^2]t)]~,
\ee
which, except for $v=1$, is the well-known 2-soliton solution
(hence the classification of this solution as a 2-soliton lattice). 
For $v=1$, however, it corresponds to the one-soliton solution.
The latter implies $k=m$ and $c=a$ from Eq. (\ref{aveq10}).

\begin{figure}
\begin{center}
\begin{tabular}{cc}
    \includegraphics[width=\middlefig]{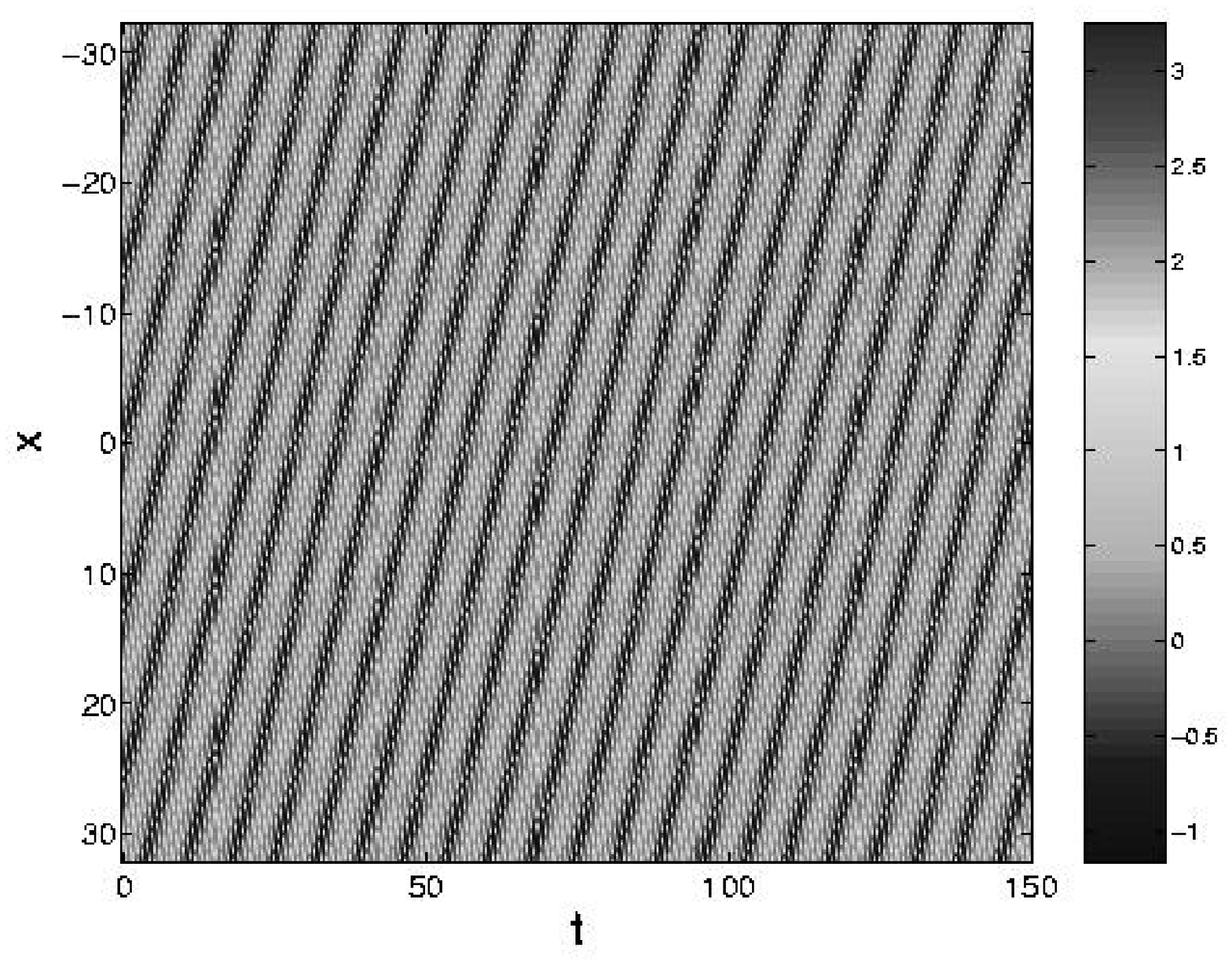}&
    \includegraphics[width=\middlefig]{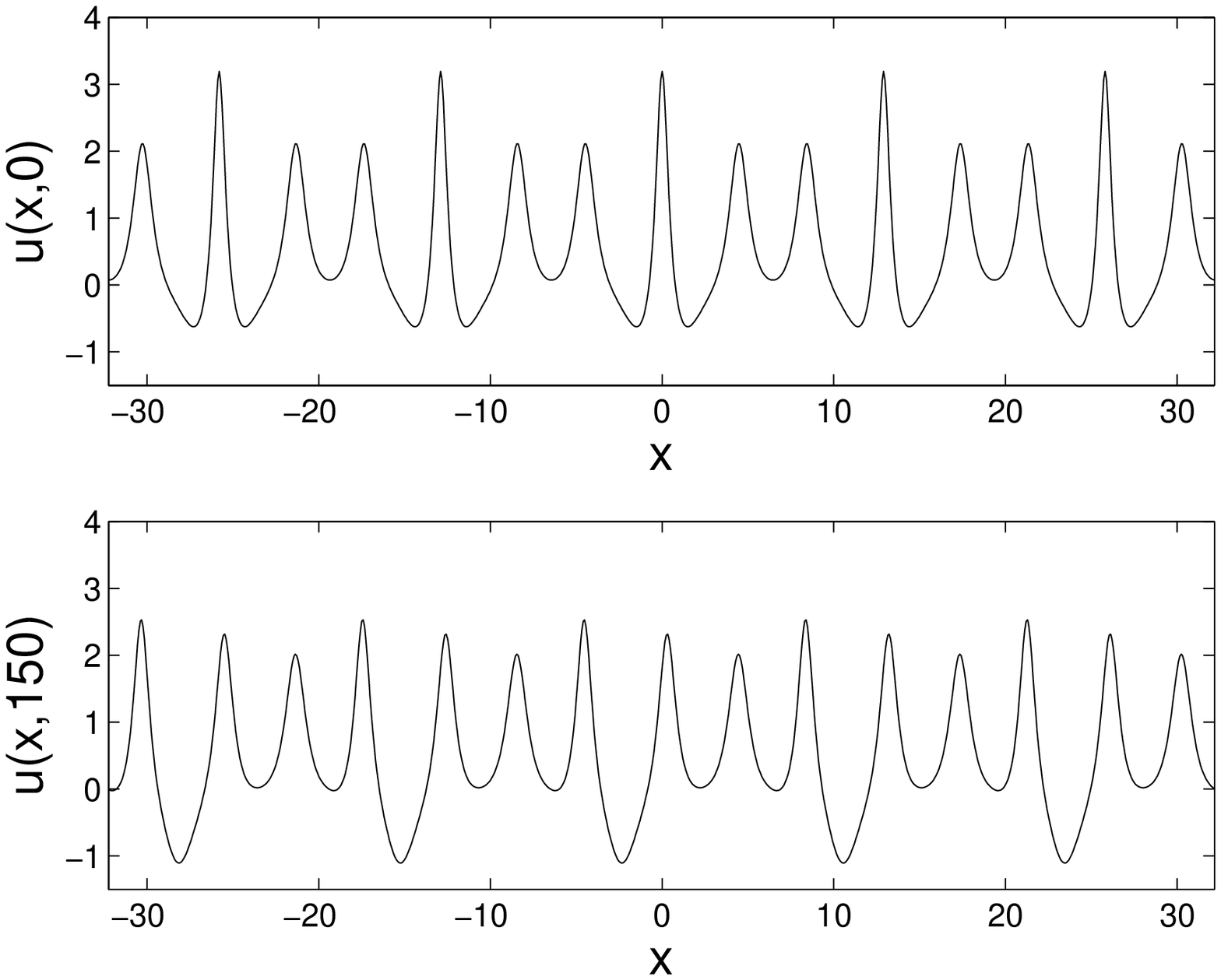}\\
\end{tabular}
\caption{The left panel shows the spatio-temporal evolution
of the contour plot of a two-soliton lattice in the case of
the positive mKdV equation for $a=0.25$, $k=0.1$, $c=1.596$ and
$m=0.999$, such that $K(k)/a=2 K(m)/c$. The right panel shows
the $t=0$ (top panel) and the $t=150$ (bottom panel) spatial profiles.}%
\label{avfig2}
\end{center}
\end{figure}

An example of the dynamical evolution for the 
two-soliton lattice solution is shown in Fig. \ref{avfig2}.
As it can be seen, this solution persists unchanged for
long dynamical evolutions
(i.e., for times of the order of 150 in the arbitrary time units
of the time evolution of Fig. \ref{avfig2}), hence our numerical
simulations indicate that it is dynamically {\it stable}.  

\section{Lattice Solutions of the Negative mKdV Equation} 

The negative mKdV equation is given as 
\begin{equation}
u_t-6u^2 u_x+u_{xxx} =0\,.
\label{aveq13}
\end{equation}
In this case, to identify the corresponding solutions,  
we start with the ansatz
\begin{equation}
u=-2\frac{d}{dx} [\tanh^{-1} \phi]\,. 
\label{aveq14}
\end{equation}
One can then show that the field $\phi$ satisfies the equation 
\begin{equation}
(1-\phi^2)[\phi_t+\phi_{xxx}]+6\phi_x[\phi \phi_{xx} -\phi_x^2]=0.
\label{aveq15}
\end{equation}
Using an ansatz in terms of Jacobi elliptic functions we obtain the following
new periodic solution:
\begin{equation}
\phi = \alpha \dn(ax+bt,k)\dn(cx+dt,m),
\label{aveq16}
\end{equation}
where 
\begin{eqnarray}
\frac{c^4}{a^4}=\frac{1-k}{1-m},~~ \frac{b}{a}
=-[a^2(2-k)+3c^2(2-m)], \nonumber\\ 
\alpha^2 = \frac{1}{\sqrt{(1-k)(1-m)}},~~   
\frac{d}{c}=-[c^2(2-m)+3a^2(2-k)].
\label{aveq17}
\end{eqnarray}
Unfortunately, this solution  may be
singular for a finite $x$ (for a given $t$), depending on the parameters. 
In particular, the derivative of Eq. (\ref{aveq14}) induces a
term $\sim 1/(1-u^2)$. However, from the properties of the
elliptic functions, one can obtain that:
\begin{eqnarray}
\sqrt{(1-k)(1-m)}<u^2<\frac{1}{\sqrt{(1-k)(1-m)}} . 
\label{aveq17a}
\end{eqnarray}
Equation (\ref{aveq17a}), in turn, implies (given the continuity of $u$) that
$u$ will, typically, assume the value $1$ for a certain $x$,
hence that the corresponding solution will, generically, be
singular. We thus do not consider it further here.


Another, more interesting solution of the negative mKdV equation is given by  
\begin{equation}
\phi = \alpha \sn(ax+bt,k)\sn(cx+dt,m),
\label{aveq19}
\end{equation}
where 
\begin{eqnarray}
 \frac{c^4}{a^4}=\frac{k}{m},~~ \frac{b}{a}=[a^2(1+k)+3c^2(1+m)], \nonumber\\ 
\alpha^2 = \sqrt{km},~~   
\frac{d}{c}=[c^2(1+m)+3a^2(1+k)].
\label{aveq20}
\end{eqnarray}
For $k=m$, this solution degenerates to the well-known soliton-lattice
solution which in the limit $k=m=1$ goes over to the famous
one-soliton solution of mKdV with negative sign.

\begin{figure}
\begin{center}
\begin{tabular}{cc}
    \includegraphics[width=\middlefig]{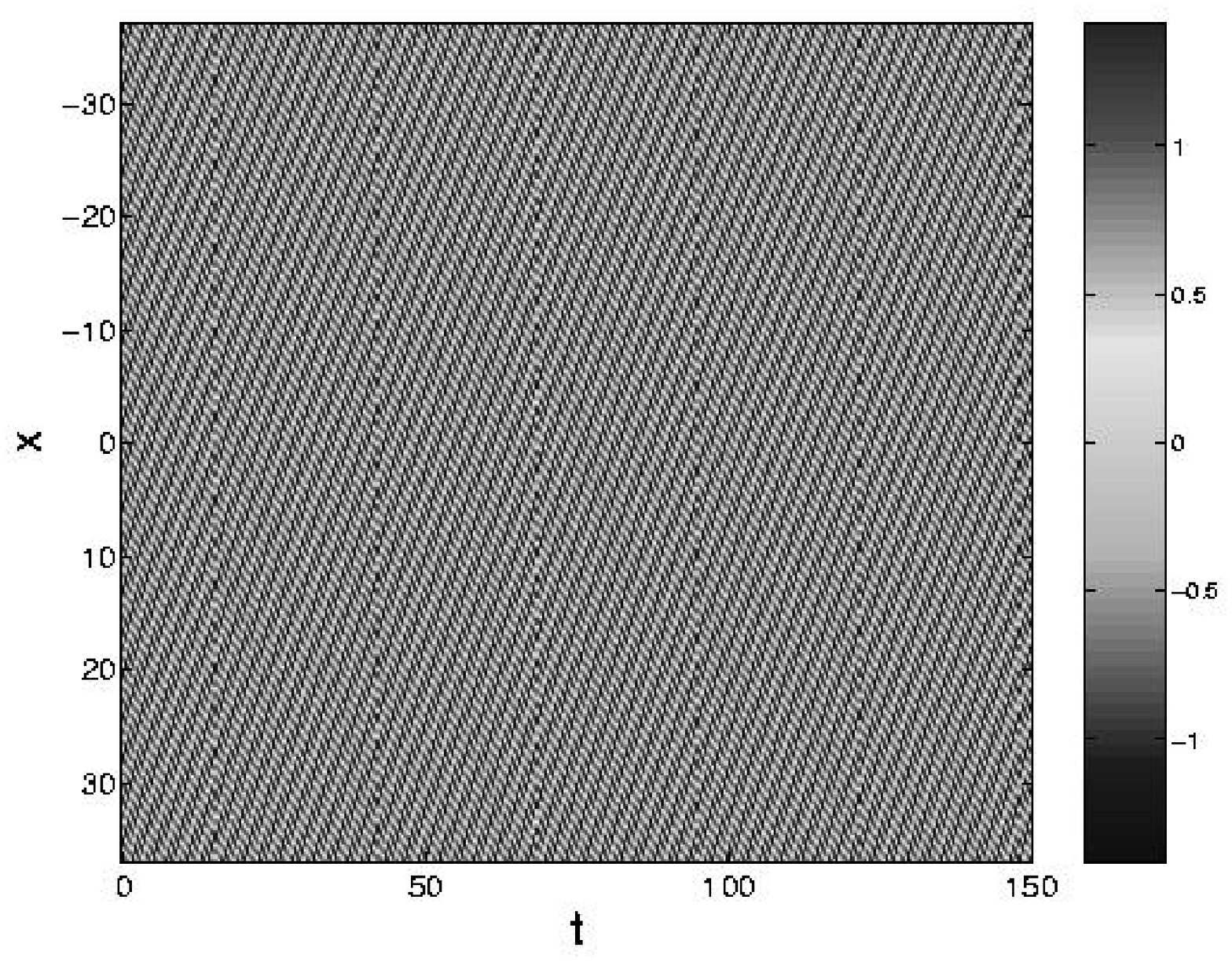} &
    \includegraphics[width=\middlefig]{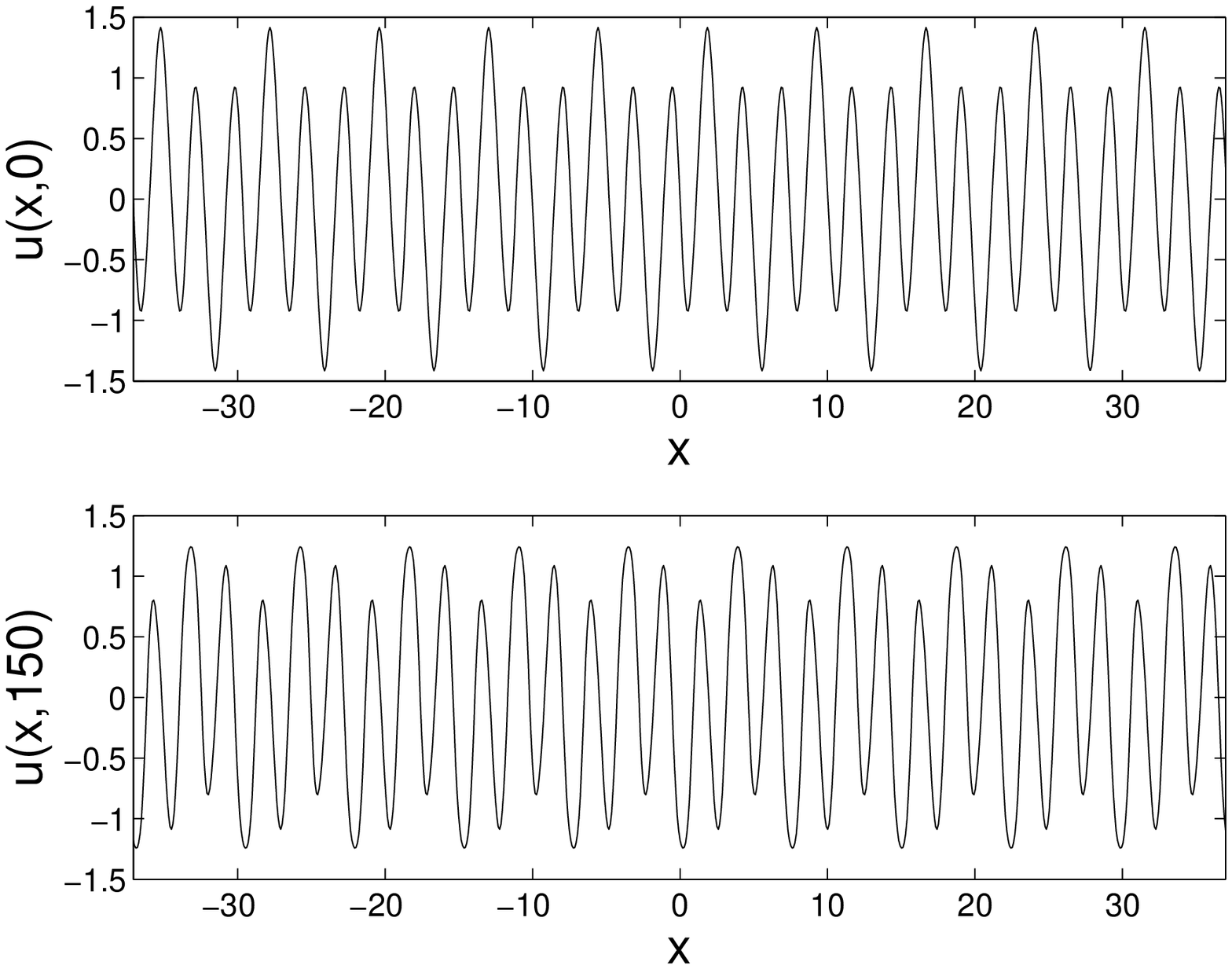}\\
\end{tabular}
\caption{Same as Fig. \ref{avfig2} but for the 
periodic solution of the negative mKdV equation with $a=1$, $k=0.5$,
$c=1.719$ and $m=0.057$ (such that $K(k)/a=2 K(m)/c$). 
The dynamical  evolution is {\it stable}. }%
\label{avfig3}
\end{center}
\end{figure}

We have examined the solution (\ref{aveq19}) 
for strong, as well as weak commensurability.
We have performed numerical simulations for 
various $(p,q)$ pairs including the degenerate case with
$c=a$ amd $k=m$. 
The principal feature of the temporal evolution of this
solution lies in its
dynamical stability, in the sense that for the duration of the
numerical simulation (of the order of $t=150$ time units) 
the structure robustly maintains its character, see Fig. 
\ref{avfig3}.  

\section{Periodic Solution of the KdV Equation} 
Through the Miura transform $v=u^2\pm u_x$ \cite{miura} we can obtain 
the periodic solutions of the KdV 
equation ($v$) from the corresponding solutions of the negative mKdV 
equation ($u$). In particular, to the solution
(\ref{aveq19}) of negative mKdV, corresponds the KdV solution of the form:
\be\label{aveq21}
v=u^2+u_{x}=2\frac{[2\phi'^{2}(x)-(1+\phi)\phi''(x)]}{(1-\phi^2)(1+\phi)}\,,
\ee
where prime denotes derivative with respect to the argument. The other
solution $u^2 -u'(x)$ is obtained from here simply by changing
$\alpha$ to -$\alpha$. Using the
solution (\ref{aveq19}) of the negative mKdV equation, we find that
the solution (\ref{aveq21}) can be expressed in the form
\bea\label{aveq22}
&&v=\frac{2\alpha}{[1+\alpha \sn(ax+bt,k) \sn(cx+dt,m)]^2}
\bigg [(a^2(1+k)+c^2(1+m)) \sn(ax+bt,k)  \nonumber \\
&&\sn(cx+dt,m) 
+2a^2 \alpha \sn^2 (cx+dt,m) +2c^2 \alpha \sn^2 (ax+bt,k) \nonumber \\
&&-2ac \cn(ax+bt,k) \dn(ax+bt,k) \cn(cx+dt,m) \dn(cx+dt,m) \bigg ]\,. 
\eea

Perhaps more interestingly, the Miura transform is an {\it exact}
transformation between the solution of the negative mKdV equation
($\forall (x,t)$) and the one of the KdV equation. This implies
that the numerically observed stability of the above mentioned 
lattice solution of the 
negative mKdV equation carries over to the existence and
stability of such a solution in the setting of the KdV equation.

\section{Conclusion} 
In this Letter, we have obtained new classes of periodic  
solutions for the positive and negative mKdV equations.  The
positive mKdV solutions (\ref{aveq7}) and (\ref{aveq9}) 
could be identified as the breather lattice and two-soliton lattice
solutions,  since in the appropriate limit they reduce to 
single breather and two-soliton solutions, respectively.  
In the case of the negative mKdV equation, 
lattice solutions have been identified in the form of Eqs. 
(\ref{aveq16}) and (\ref{aveq19}). However, the latter have not
been designated as breather or soliton lattices (as the process of
obtaining limiting expressions is less straightforward for the 
negative mKdV case).

We have also examined the stability of the various periodic solutions of
mKdV equation with both positive and negative signs.  
We have found that the two-soliton lattice solution of the positive mKdV
and the lattice solution (\ref{aveq19}) of the negative mKdV 
are quite robust with respect to perturbations in contrast with the 
breather lattice solution of the positive mKdV equation.  
These results, and more specifically the stability of the negative
mKdV lattice solution, 
have direct 
implications for the corresponding lattice solutions of the KdV equation. 

This work was supported in part by the U.S. Department of Energy. 
PGK is grateful to the Eppley Foundation
for Research, the NSF-DMS-0204585 and the NSF-CAREER program for 
financial support.



\vspace{5mm}


\section*{References}

\begin{thebibliography}{99}

\vspace{2mm}


\bibitem{mkdv1} P.G. Drazin and R.S. Johnson, 
\newblock {\it Solitons: an introduction }
\newblock (Cambridge University Press, Cambridge, U.K., 1989).  

\bibitem{AS} M.J. Ablowitz and H. Segur,
\newblock {\it Solitons and the Inverse Scattering Transform}
\newblock (SIAM, Philadelphia, 1981).

\bibitem{IR} E. Infeld and G. Rowlands,
\newblock {\it Nonlinear Waves, Solitons and Chaos}
(Cambridge University Press, Cambridge, 1990).

\bibitem{Dodd} R.K. Dodd, J.C. Eilbeck, J.D. Gibbon and H.C. Morris,
\newblock {\it Solitons and Nonlinear Waves} (Academic Press, London, 1982).

\bibitem{Scott} A. Scott, 
\newblock {\it Nonlinear Science} (Oxford University Press, New York, 1999). 

\bibitem{ion} K. E. Lonngren, Optical and Quantum Electronics, 
{\bf 30}, 615 (1998). 

\bibitem{plasma} A. H. Khater, O. H. El-Kalaawy, and D. K. Callebaut, 
Phys. Script. {\bf 58}, 545 (1998).  

\bibitem{schot} V. Ziegler, J. Dinkel, C. Setzer, and K. E. Lonngren, 
Chaos, Solitons and Fractals {\bf 12}, 1719 (2001).  
 
\bibitem{traffic} T. S. Komatsu and S. I. Sasa, Phys. Rev. E {\bf 52},
5574 (1995); T. Nagatani, Physica A {\bf 265}, 297 (1999). 

\bibitem{ono} H. Ono, J. Phys. Soc. Jpn. {\bf 61}, 4336 (1992). 

\bibitem{schief} W. K. Schief, Nonlinearity {\bf 8}, 1 (1995). 

\bibitem{slag} M. Agop and V. Cojocaru,
Mater. Trans. JIM {\bf 39}, 668 (1998). 


\bibitem{jets} E. A. Ralph and L. Pratt, J. Nonlin. Sci. {\bf 4}, 
355 (1994). 


\bibitem{mat} S. Matsutani and H. Tsuru, J. Phys. Soc. Jpn. {\bf 60}
(1991) 3640.

\bibitem{gol} R.E. Goldstein and D.M. Petrich, Phys. Rev. Lett. {\bf 67},
3203 (1991); J. Langer and R. Perline, Phys. Lett. A {\bf 239}, 
36 (1998); K. S. Chou and C. Z. Qu, Physica D {\bf 162}, 9 (2002). 


\bibitem{sg} P.G. Kevrekidis, A. Saxena, and A.R. Bishop, Phys. 
Rev. E {\bf 64}, 026613 (2001). 

\bibitem{mkdv} P.G. Kevrekidis, A. Khare, and A. Saxena, Phys. Rev. 
E {\bf 68}, 047701 (2003). 

\bibitem{miura} R. Miura, J. Math. Phys. {\bf 9}, 1202 (1968).

\bibitem{ohta} Y. Ohta and R. Hirota,
\newblock J. Phys. Soc. Jpn. {\bf 60}, 2095 (1991).

\bibitem{ladik} M.J. Ablowitz and J.F. Ladik,
\newblock J. Math. Phys., {\bf 16}, 598 (1975);
M.J. Ablowitz and J.F. Ladik,
\newblock J. Math. Phys., {\bf 17}, 1011 (1976).


\end{thebibliography}
\end{document}